\documentclass[twocolumn,showpacs,preprintnumbers,amsmath,amssymb]{revtex4}
\usepackage{graphicx}
\begin{document}
\def \be{\begin{equation}}
\def \ee{\end{equation}}
\def \bea{\begin{eqnarray}}
\def \eea{\end{eqnarray}}
\def \half{{1\over 2}}
\def \etal{{\it et~ al~}}
\def \cH{{\cal{H}}}
\def \cE{{\cal{E}}}
\def \cD{{\cal{D}}}
\def \cM{{\cal{M}}}
\def \cN{{\cal{N}}}
\def \bS{{\bf S}}
\def \bl{\mbox{\boldmath{$l$}}}
\def \bx{{\bf x}}
\def \bq{{\bf q}}
\def \bfe{{\bf e}}
\def \bL{{\bf L}}
\def \e{{\epsilon}}
\def \a{{\alpha}}
\def \t{{\theta}}
\def \b{{\beta}}
\def \g{{\gamma}}
\def \D{{\Delta}}
\def \d{{\delta}}
\def \w{{\omega}}
\def \s{{\sigma}}
\def \nd{{^{\vphantom{\dagger}}}}
\def \yd{^\dagger}
\def \ket#1{{\,|\,#1\,\rangle\,}}
\def \bra#1{{\,\langle\,#1\,|\,}}
\def \braket#1#2{{\,\langle\,#1\,|\,#2\,\rangle\,}}
\def \expect#1#2#3{{\,\langle\,#1\,|\,#2\,|\,#3\,\rangle\,}}
\def \rl#1#2{{\,\langle\,#1\,#2\,\rangle\,}}
\def \prb{{Phys. Rev. B}}

\title{Low Energy Singlets in the Heisenberg Kagome Antiferromagnet}
 \author{Ran Budnik$^1$ and Assa Auerbach$^{1,2}$\\
 $^1$Physics Department, Technion, Haifa 32000, Israel.\\
$^2$The Physics Laboratories, Harvard University, Cambridge MA
02138.}
\date{\today}
\begin{abstract}
The spin half Heisenberg antiferromagnet on the Kagom\'e lattice,
is mapped by Contractor Renormalization to a Spin-Pseudospin
Hamiltonian on the triangular superlattice. Variationally, we find
a ground state with columnar dimer order. Dimer orientation
fluctuations are described by an effective O(2) model at energies
above an exponentially suppressed clock mass scale. Our results
explain the large density of low energy singlets observed
numerically, and the non magnetic $T^2$ specific heat observed
experimentally.
\end{abstract} \maketitle
\narrowtext Frustrated quantum antiferromagnets (AFM) are
important paradigms of emerging phenomena in models of condensed
matter. It has long been appreciated that due to the extensive
{\em classical} ground state manifold of the Kagom\'e
lattice\cite{kagome-veit}, quantum fluctuations may destroy
magnetic order and  replace it with paramagnetic phases with novel
excitations at low energy scales.

Numerical studies of the spin half Kagom\'e AFM have suggested
that its ground state does not support long range spin
order\cite{ZE,CE,numerics-review}. The low spectrum of the
Kagom\'e \cite{singlets} consists of {\em singlets} whose number
increases exponentially with the lattice size.

Experimentally, there is amounting evidence of unusual low energy
excitations in Kagom\'e like systems. In spin-$\frac{3}{2}$
$SrCr_9Ga_{12}O_{19}$\cite{neutrons,exp1} a significant fraction
of the spin moment is {\em not frozen} below the non linear
susceptibility maximum at T=5K. Recently, muon resonance
experiments on a spin-$\frac{1}{2}$ system\cite{exp-spinhalf},
reported that below the susceptibility maximum of T=20K, low
frequency spin fluctuations were detected but {\em without static
magnetization} down to 50mK.

However, the specific heat of  $SrCr_9Ga_{12}O_{19}$  has an
unexplained large $T^2$ coefficient which apparently does not
arise from spin waves\cite{exp1}.

Thus there is both numerical and experimental evidence that there
are seemingly {\em non magnetic} massless modes whose origin has
not yet been understood. Weak bonds perturbation theory has been
applied\cite{PT,MM}, and interesting results have been found for
the Quantum Dimer Model (QDM)\cite{QDM,numerics-review} on the
Kagom\'e lattice. However, the QDM has not yet been quantitatively
derived from the Heisenberg model.
\begin{figure}
\begin{center}
\includegraphics[width=5.5cm]{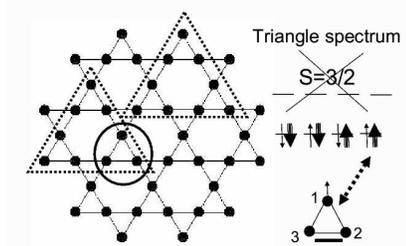}
\caption {CORE on the Kagom\'e lattice (solid circles). Triangular
blocks of first (second) CORE steps are encircled by solid
(dashed) lines. On the right: triangle four ground states are
labelled by spin (arrows) and pseudospins (wide arrows). The
$\uparrow\Uparrow$ state of the triangle is visualized as a dimer
singlet on the bottom rung.} \label{fig:coresteps}
\end{center}
\end{figure}

It is the purpose of this Letter to elucidate the nature of low
energy excitations of the S=1/2 Kagome AFM by applying the
Contractor Renormalization (CORE) method\cite{MW},  which is {\em
not perturbative} in weakened bonds. CORE has been recently
applied to the square lattice Hubbard model\cite{AA}, Heisenberg
ladders\cite{PS} (including detailed convergence tests\cite{CP}),
and the frustrated Checkerboard and Pyrochlore lattices\cite{BAA}.

For the Kagom\'e lattice, CORE leads to an effective
Spin-Pseudospin (S-L) model on a triangular lattice. A variational
analysis reveals {\em columnar dimer order} in the spin disordered
ground state, and low energy excitations which can be understood
as dimer orientation fluctuations. We describe these fluctuations
by a $p=6$ Quantum Clock Model. Its  mass gap is strongly
suppressed  by quantum fluctuations by a factor estimated at about
$10^{-4}$. Thus, the low energy singlet spectrum is in effect a
quasi-Goldstone mode of an O(2) order parameter. The number of sub
gap singlets increases exponentially with lattice size and gives
rise to a $T^2$ specific heat as seen experimentally.

The Heisenberg Hamiltonian on the Kagom\'e lattice, (see Fig.
\ref{fig:coresteps}),  is
\begin{equation}
H=J\sum_{\langle ij\rangle} \bS_i\cdot\bS_j\label{eq:heisenberg},
\end{equation}
Henceforth we set the unit of energy to $J=1$.

{\em Blocks.} CORE involves (i) an initial choice of elementary
blocks which cover the lattice, and (ii) a truncated set of block
eigenstates whose tensor product spans the reduced Hilbert space.
It is useful to choose minimally sized blocks which respect (as
much as possible) the lattice point group symmetry. Here we select
upward triangles, and a truncated basis of the four  degenerate
spin half ground states, discarding the higher $S=3/2$ states,
(see Fig.(\ref{fig:coresteps})).

The S-L representation of the four ground states are labelled by
$|s,l\rangle$, where $s=\uparrow,\downarrow$ is the magnetization
and $l=\Uparrow,\Downarrow$ is the pseudospin in the $z$
direction. Explicitly, in the Ising basis $|s_1 s_2 s_3\rangle $,
 \bea
|s, \Uparrow \rangle &=&    {(|s\uparrow \downarrow \rangle
 -|s\downarrow \uparrow \rangle ) \over \sqrt{2}}\nonumber\\
|s, \Downarrow \rangle  &=& {|s \uparrow \downarrow \rangle
+|s\downarrow \uparrow \rangle ) \over \sqrt{6}} - \sqrt{2\over 3}
| (-s)  s  s \rangle  \label{eq:states} \eea

The pseudospin direction in the $xz$ plane correlates with the
direction of the singlet bond, e.g. $\Uparrow$ describes a singlet
dimer on the bottom ($-\hat{z}$) edge (see
Fig.\ref{fig:coresteps}). Thus, the $L^y$ eigenstates have
definite chiralities.

{\em Effective Hamiltonian}. The effective interactions between
triangles is calculated by CORE\cite{MW,AA}. We note that this
approach is feasible  when two conditions are met: (i) Interaction
matrix elements  fall off rapidly with range such that the
truncation error at finite ranges is small, and (ii) the norms of
the projected eigenstates are sufficiently large for numerical
accuracy.  We have computed all range 2 and range 3 interactions,
and neglected range 4 corrections, whose expectation values were
found to be an order of magnitude smaller. At range 3, norms of
projected eigenstates were greater than 0.75, with most states
above 0.9.

\begin{figure}[ht]
  \centering
  \includegraphics[height=3.8cm,width=6cm]{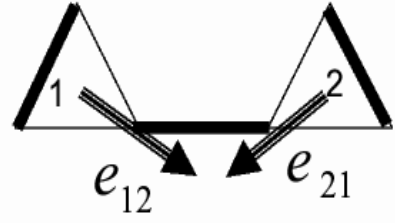}
  \caption
  {CORE range 2. Directions of anisotropy vectors
$\bfe_{ij}=\bfe^l_{ij}= \tilde{\bfe}_{ij} $
  for a horizontal bond. For other bond directions, vectors must be rotated by $\pm
  120^\circ$. The ground state singlet correlations
  of two  coupled blocks are depicted by thick lines.
  }
  \label{fig:directions}
\end{figure}

The effective Hamiltonian is a Spin-Pseudospin (SL) Model on the
triangular lattice:
\begin{eqnarray}
\cH_{SL}&=&\cH_{ss}+\cH_{ll}\nonumber\\
\cH_{ss}&=&\sum_{\langle ij\rangle} \bS_i \cdot \bS_j~
\left[J_{ss}+J_{sslele}( \bL_i\cdot {\bf e}_{ij} ) \cdot (
\bL_j\cdot{\bf e}_{ji} )
\right.   \nonumber\\
& & +J_{ssll}(\bL^\perp_i\cdot \bL^\perp_j)+ J_{ssle1}(\bL_i\cdot{\bf e}_{ij} ) \nonumber\\
&& \left.+J_{ssle2}(\bL_j\cdot{\bf e}_{ji} )+J_{sslyly}\bL_i^y \bL_j^y    \right]  \nonumber\\
\cH_{ll}&=&J_{lele}( \bL_i\cdot {\bf \tilde{e}}^{ij} ) \cdot (
\bL_j\cdot{\bf \tilde{e}}_{ji} )+
J_{ll}(\bL^\perp_i\cdot \bL^\perp_j)  \nonumber\\
&&+ J_{lyly} \bL_i^y \bL_j^y   \label{Heff}
\end{eqnarray}
Here $\bL^\perp=(\bL^x,\bL^z)$, and $\bfe_{ij},\bfe_{ij}^l$ are
unit vectors in the $xz$ plane. $\cH_{ss}$ describes interactions
of the Kugel-Khomskii type\cite{KK,MM}, where the pseudospin
exchange anisotropy depends on the sites and bond directions. (See
Fig.\ref{fig:directions}). For any other other bond $\langle
ij'\rangle$,  $\bfe_{ij'}$ is simply found by rotating $\bfe_{ij}$
by $0,\pm 120^{\circ}$ according to the O(2) rotation of $ \langle
ij \rangle \to \langle ij'\rangle$.

The coupling constants and angles of $\cH_{SL}$,  are tabulated in
Table \ref{tab:1}. Missing from  (\ref{Heff}) are terms which
vanish in the periodic lattice by summation over nearest
neighbors, and three site (ring exchange) interactions.  The
largest term $K \bS_i \cdot \bS_j L^z_i L^z_j L^z_k$ has a largest
matrix element of magnitude $0.02$.
\begin{table}
\begin{tabular}{|c c c c c c c c|} \hline
 $J_{ss}$&$J_{sslele}$ & $J_{ssll}$ &  \underbar{$J_{ssle_1}$}&
 \underbar{$J_{ssle_2}$}&$J_{sslyly}$ &$\bfe_{12}$& $\bfe_{21}$ \\
 0.108& 0.954&0.211 & 0.281& 0.278& 0.053 &  113$^\circ$&  248$^\circ$  \\\hline
$J_{lele}$ & $J_{ll}$&$J_{lyly}$& $\bfe^l_{12}$& $\bfe^l_{12}$ & & & \\
0.060& -0.001&0.038 & 132$^\circ$ & 222$^\circ$&&&\\\hline
\end{tabular}
\caption{CORE up to range 3: Interaction parameters for the
effective hamiltonian $\cH_{SL}$, Eq. (\ref{Heff}). Underlined are
the ''Dimerization Fields'' (see text). }\label{tab:1}
\end{table}

While $\cH_{SL}$ may prove to be useful for numerical studies of
the spectrum of larger lattices,  its complexity somewhat obscures
its physics. It is simple however to study $\cH_{SL}$
variationally using pseudospin coherent states\cite{book}
$\psi[s]\prod_i |s_i,\bl_i\rangle$ where $\bL \cdot \bl|s, \bl
\rangle=\half|s, \bl \rangle$. Its energy is minimized with
respect to the directions $\bl_i$.

\be E^{var}[\bl]=\sum_{ij} \langle \bS_i \bS_j\rangle_{[\bl]}
F[\bl]+E_{ll}[\bl] \ee

We start by evaluating the energy of the magnetically ordered
state, where both the spins and the pseudospins form three
sublattice (3SL)  N\'eel order on the triangular lattice (and
$\sqrt{3}\times\sqrt{3}$ order on the Kagom\'e). Other candidates
are the dimer coverings of two triangle singlets, whose
correlations are defined by Fig.~\ref{fig:directions}. The dimer
singlet states have been shown by Mila and Mambrini\cite{MM} to
span much of the low singlet spectrum in finite cluster
calculations. The variational analysis highlights the special role
of the {\em ``Dimerization Fields''}, $J_{ssle_1},J_{ssle_1}$ in
(\ref{Heff}), for the formation of local singlets. These terms
cancel under summation in all uniform states defined by $\langle
\bS_i \bS_j\rangle=\mbox{const}$. Their significant magnitude (see
Table~\ref{tab:1}) helps to lower the energy considerably by
aligning $\bl_i$ with the anisotropy vectors $\bfe_{ij}$ to form
singlets on certain bonds and not others $\langle \bS_i \cdot
\bS_j \rangle=-{3\over 4}\delta_{\langle ij\rangle_{dim} }$. {\em
This is a strong argument in favor of a paramagnetic ground
state.} Consequently, $\cH_{ll}$ is crucial in selecting the true
ground state among the multitude of dimer singlet coverings. We
have found that the perfectly ordered {\em columnar dimer} (CD)
state minimizes $\cH_{ll}$.  A local ``defect'' of a rotated dimer
in the CD background costs a ``twist'' energy of $+0.01$ per site.

\begin{figure}
\begin{center}
\includegraphics[width=6cm]{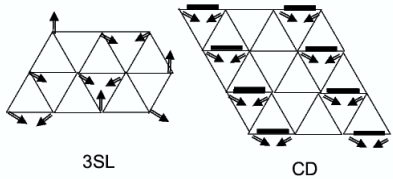}
\caption {Variational states on the triangular superlattice: The
Three Sublattice (3SL)  and the Columnar Dimer (CD) State, arrows
denote pseudospins, and thick lines denote singlet dimers.}
\label{fig:variational}
\end{center}
\end{figure}

In Fig. \ref{fig:variational} we depict the  3SL and CD states.
Their energies per site are \be E_{CD}/\mbox{site} =
-0.229,~~~~~~~E_{3SL}/\mbox{site} = -0.178,
\label{eq:var-energies}. \ee
where the evaluation uses the known
spin correlations of the Heisenberg AFM on the triangular
lattice\cite{capriotti-rev} $\langle \bS_i \bS_j\rangle = 0.18$.

This result can be connected to  numerical diagonalizations data
as follows. The number of CD states on  Kagome clusters with the
lattice group symmetry is 24: From a particular up-triangle site
there are 6 dimer directions. There are two equivalent dimer
orderings of the neighboring lines of dimers. Another factor of
two is given by the down triangle configurations. Between these 24
CD states there are exponentially vanishing overlaps at large
lattices.

\begin{figure}
\begin{center}
\includegraphics[width=5.5cm]{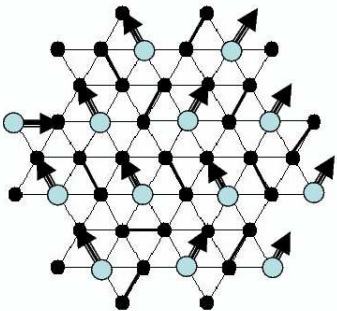}
\caption {Clock fields $\phi(\bx)$ (thick arrows) defined by the
dimer directions from the sites marked by large circles.}
\label{fig:clock}
\end{center}
\end{figure}

{\em Quantum Clock Model.} A continuum Hamiltonian for the low
energy fluctuations is derived as follows. Using the subset of
site positions $\bx$ belonging to a $2\times 2$ superlattice (see
Fig. \ref{fig:clock}), every dimer configuration defines a unique
configuration of 6-fold clock angles $\phi(\bx)$'s defining the
orientations of dimers from the selected sites.

A ferromagnetic state of $\phi$'s represents a CD ground state (up
to global translations, with vanishing overlap, of an
interpolating line of dimers). The resistance to local twists,
governed by $J_{lele}$, is described by energy density
$\half\rho_s(\nabla\phi)^2, \rho_s \simeq 0.01$. The barrier
height between dimer orientations is estimated from
Eq.(\ref{eq:var-energies}) to be $h_6=0.05$ which defines the
``clock field''  $h_6\cos(6\phi)$.  $J_{lyly} L_yL_y$ interactions
rotate the pseudospins and the dimers in the plane, giving rise to
a kinetic energy $\half \chi \dot{\phi}^2, 1/\chi \simeq 0.01$.

Thus we arrive at a long wavelength partition function of a (2+1)D
Quantum Clock Model (QCM) which describes the lowest singlet
sector of the Kagom\'e spectrum:
\bea Z_{\mbox{singlets}}&=&\int
\cD \phi
\exp\left( - \int_0^\beta d\tau d^2\bx ~\right.\nonumber\\
&&~~~\left. \left( \half\chi \dot{\phi}^2 +\half\rho_s (\nabla
\phi)^2 + h_6 \cos(6\phi ) \right)\right) \label{QCM} \eea

The renormalization group analysis of Jos\'e {\em et.
al.}\cite{Jose} for the {\em classical} $p$-state Clock Model in
two dimensions found that at low temperatures  $T < T_p=8\pi/p^2$
the clock field $h_p$ is a relevant interaction which locks the
ground state into a clock minimum, with a finite gap for domain
wall excitations. For our QCM  the clock term is still relevant,
but quantum fluctuations, which involve tunnelling between clock
minima, drastically renormalize down the value of the clock field
and the mass gap for the long wavelength excitations.

This is shown as follows: we expand the action of (\ref{QCM}) to
lowest order in $h_6\cos(6\phi)$ and integrate out the high
momenta and energy modes $\phi_{>}$ \bea
 h_6 \int &\cD&\phi_{\tiny >}  \cos(p (\phi_{\tiny <}+\phi_{\tiny >}) )
 \exp\left( -\half \phi_{\tiny > }G^{-1} \phi_{\mbox{\tiny >}}  \right)
\nonumber\\
&=& h_6 \cos(p\phi_{\tiny <} )
e^{-p^2 \langle \phi_{>}^2 \rangle },\nonumber\\
h_6^{ren}&=& h_6 e^{-C p^2  }\approx 0.05\cdot 10^{-4}
\label{eq:h6} \eea

Our rough numerical evaluation of  $C\approx 0.2$ uses the zero
point phase fluctuations of an effective spin 2 quantum $xy$
model, describing  four pseudospin  half in the unit cell. Eq.
(\ref{eq:h6}) is our key result: tunnelling between $p=6$ ground
states renormalizes down the clock mass gap by a gaussian function
of $p$. In particular, the dispersion of $\phi$ fluctuations
 \be
\omega^{S=0}_\bq = \sqrt{(h^{ren}_6)^2 + {\rho_s\over
\chi}|\bq|^2} \label{omegaq} \ee resembles Goldstone modes of an
O(2) $xy$-model at frequencies and temperatures larger than
(\ref{eq:h6}). In this regime, the singlets' contribution to the
specific heat is quite large. Using thermodynamics of free bosons,
we obtain

\be C_V \sim {\chi  \over \rho_s} T^2,~~~S (E) \sim N
\left({E\over N 0.01}\right)^{2/3},
 \ee
where $N$ is the number of effective sites. Thus, the smallness of
$h^{ren}_6$ provides the long sought after explanation of the
unusual exponential proliferation of singlets at sub gap
energies\cite{singlets}. In addition, the singlets
pseudo-goldstone mode can now explain the experimentally reported
$T^2$ term in the specific heat\cite{exp1}.

We note that the CD state has no long range spin order.  The spin
gap can be estimated from the variational energy difference
between CD and 3SL states to be of the order of $h_6=0.05$.
Another estimate can be obtained by iterating CORE on $H_{SL}$.

At the second CORE step, the triangular lattice is covered by
triangles (which are blocks of 9 Kagom\'e lattice sites). These
form a triangular superlattice with directed bonds (see
Fig.\ref{fig:coresteps}). For both the Heisenberg and $\cH_{SL}$,
each block has four degenerate $S=1/2$ ground states which can
again be represented by a spin and a pseudospin. The second CORE
step thus maps the S-L Hamiltonian (\ref{Heff}) onto itself with
new interaction parameters and anisotropy vectors,  as listed in
Table \ref{tab:2}.

\begin{table}
\begin{tabular}{|c c c c c c c c|} \hline
 $J_{ss}$&$J_{sslele}$ & $J_{ssll}$ &  $J_{ssle_1}$ &
  $J_{ssle_2 }$ &$J_{sslyly}$ &$\bfe_{12}$& $\bfe_{21}$ \\
 0.113& 0.08& -0.005 & 0.026& 0.182& -0.039 &  330$^\circ$&  280$^\circ$  \\\hline
$J_{lele}$ & $J_{ll}$&$J_{lyly}$& $\bfe^l_{12}$& $\bfe^l_{12}$ & & & \\
-0.019& -0.003&0.004& 200$^\circ$ & 160$^\circ$&&&\\\hline
\end{tabular}
\caption{Second CORE iteration: Interaction parameters  for
$\cH_{SL}$ evaluated up to range 2. }\label{tab:2}
\end{table}
{\em What can we learn from step 2?}

In contrast to the first CORE step (see Table \ref{tab:1}), the
vectors $\bfe_{ij}\cdot\bfe_{ji}>0$, i.e.  ferromagnetic.
$J_{sslele}$ has decreased while $J_{ss}$ did not. Thus the
Hamiltonian prefers local singlet correlations with aligned
pseudospins, which is consistent with columnar order.

{\em Spin gap.} By iterating CORE we can compute the splitting
between the $S=\half$ ground state and the lowest $S={3 \over 2}$
excitation on triangular clusters. In Table~\ref{tab:spingap} the
spin gap is computed on up to 81 original Kagom\'e lattice sites.
At third CORE step, many wave function overlaps vanish. This is
expected due to the onset of long range CD order, since states
with high pseudospin parentage have lower energy.  Stopping at 81
sites, we can only observe that while the spin gap is larger than
the dimer fluctuations bandwidth, there is no conclusive sign of
saturation to a finite thermodynamic limit.

\begin{table}
\begin{tabular}{|c c c c c |} \hline
 lattice size& 3 & 9  &  27 & 81  \\\hline
spin gap & 1&0.5& 0.1& 0.06  \\\hline
\end{tabular}
\caption{Spin gap as a function of effective Kagom\'e lattice size
using the Heisenberg model, first and second iterations of CORE.
}\label{tab:spingap}
\end{table}

In summary, we have computed the effective Hamiltonians of the
S=1/2 Heisenberg antiferromagnet on the Kagom\'e lattices using
CORE up to range 3. Variationally, we conclude that the Kagom\'e
lattice has long range columnar dimer order and a very low gap for
singlet excitations in the thermodynamic limit.  CD order might
induce experimentally detectable lattice distortions. Further
details can be found elsewhere\cite{thesis}.

{\em Note added:}  In a recent preprint, Capponi, Laeuchli, and
Mambrini\cite{CLM} have independently computed $\cH_{SL}$ of Eq.
(\ref{Heff}) by CORE, and found excellent numerical agreement with
large cluster diagonalizations.

{\em Acknowledgements.} Discussions with E. Altman, E. Demler, C.
Lhullier and  D. Nelson are gratefully acknowledged. A.A. is
thankful for the hospitality of Harvard University, Aspen Center
for Physics, and Kavli Institute for Theoretical Physics where
part of this work has been completed. This work was supported in
part by the US-Israel Binational Science Foundation.

\end{document}